# LLMDR: Large language model driven framework for missing data recovery in mixed data under low resource regime

Durga Keshav[1], GVD Praneeth[1], Chetan Kumar Patruni[1], Vivek Yelleti[1,*], U Sai Ram[2]

[1]Department of Computer Science & Engineering, SRM University AP, India

[2]Department of Information Technology, Chaitanya Bharathi Institute of Technology, Hyderabad, India

durgakesav_tipparaju@srmap.edu.in, nagavenkatadurgapraneeth_gamidi@srmap.edu.in, chetankumar_patruni@srmap.edu.in, **vivek.yelleti@gmail.com**, **sairam.cbit@gmail.com**

## Abstract

The missing data problem is one of the important issues to address for achieving data quality. While imputation-based methods are designed to achieve data completeness, their efficacy is observed to be diminishing as and when there is increasing in the missingness percentage. Further, extant approaches often struggle to handle mixed-type datasets, typically supporting either numerical and/or categorical data. In this work, we propose LLMDR, automatic data recovery framework which operates in two stage approach, wherein the Stage-I: DBSCAN clustering algorithm is employed to select the most representative samples and in the Stage-II: Multi-LLMs are employed for data recovery considering the local and global representative samples; Later, this framework invokes the consensus algorithm for recommending a more accurate value based on other LLMs of local and global effective samples. Experimental results demonstrate that proposed framework works effectively on various mixed datasets in terms of Accuracy, KS-Statistic, SMAPE, and MSE. Further, we have also shown the advantage of the consensus mechanism for final recommendation in mixed-type data.

## 1. Introduction

Qualitative data is essential for training machine learning or deep learning models, which are built on artificial intelligence (AI) for reliable performance. The loss of data or missing data is a most critical problem in maintaining data quality due to various reasons, such as device failure or device malfunction, data collection issues, etc. To address these issues, missing value imputation methods are introduced, aiming to replace missing values using observed data samples. There are three main categorizations of missingness mechanisms [1], such as missing completely at random, Missing at random, and Missing not at random.

Existing methods support addressing the missing data using either deletion-based approaches or imputation-based approaches. In addition, existing solutions address missing values in numerical and categorical data, but not mixed data. Supporting various datatypes or mixed data types is essential in the real world, as many of the domains, such as healthcare and e-commerce, often have a mixture of different datatypes, such as text, numerical, and categorical. To maintain the data quality, different imputation methods have been introduced [2]. Some of the old methods, such as mean and mode, are rule-based methods, and K Nearest Neighbors (KNN) is used for a similarity-based method. These statistical methods [3] or KNN methods [4] have more generalization due to the applicability of rules on various datasets. Even though the generalization capacity is higher, the said methods struggle in capturing the hidden patterns in diverse data. There are various methods introduced to discover the relationship between samples and features in machine learning and deep learning, such as MICE [5], auto encoder [6], and graph neural networks[7]. These methods could handle numerical and categorical data by considering individual features effectively, but struggle with text data.

---
[*] Corresponding Author: vivek.yelleti@gmail.com

Traditional machine learning models need more preprocessing and different methods to handle missing values in various data types, such as data cleaning, imputation, feature engineering, and encoding [8]. Large language models (LLMs) have significantly shown enhanced performance by training on various data without substantial changes in their architecture. LLMs can understand the patterns that exist in the available data and are still able to produce outputs even with missing values. LLMs can learn context from the available data instead of depending on the exact values. To make better interpretations, LLMs consider the information to find the relationships between features. For example, in a dataset, if we have only a text description but have some missing values in a numeric feature, an LLM can still get the overall meaning. This enable usage of LLMs for handling real-world datasets that frequently consist of missing values in various data types. Imputation under LLM methods is classified into two main categories: finetuning methods and context methods using well-defined prompts [9].

In the literature there, to the best of our knowledge, only one work [42] reported where LLMs are employed for data recovery for tabular datasets. However, the performance of the work [42] has the following limitations:

(i) **Limited to numerical data:** Applicable over only numerical due to insufficient exploration of tabular data because LLMs need to consider the cell value and the complete table data. Because the relationships in the table may not always consider the near or pairwise values. Due to attention-based, LLMs majorly focus on the relationship between pairs of values. Therefore, LLMs that are based on sequential input data may struggle with specific relationships between columns and within columns. To address this mixed type data, which is very prominent in real-time applications with a higher accuracy, individually detecting the best suitable methods can be employed, such as MICE for numerical type of data and jellyfish [10] for text data. However, separately employing a particular method may avoid intricate dependency and reduce the overall performance.

(ii) **Computationally Expensive:** Framework of [42] considers the entire dataset whenever it recommends the missing value which is often time consuming process.

To address this, we propose an LLM framework that can address the missing data in mixed data types. The main contributions of this work are:

- Develop and design the data recovery framework for mixed data types.
- Develop a mechanism for recovering the data which could work in low-regime resources.
- Design robust mechanism which shows the stable performance irrespective of missingness ratio.

The rest of the paper is organized as follows: In Section 2, we cover the related works. Methodology is presented in Section 3, and results are presented in Section 4. The conclusion is presented in section 5.

## 2. Related Work

In statistics and machine learning, the problem of missing data and data imputation has been studied extensively. Traditional methods, such as mean, median, most frequent value, and constant values such as zero, minimum, and maximum values, are used to fill the missing values, but this method introduces bias into the dataset [11]. To overcome this limitation, other techniques such as k Nearest Neighbours (kNN) have been developed [12]. Machine learning and deep learning techniques focused on imputing values using matrix factorization [13] and auto encoder methods [14]. However, selecting the best imputation method is still challenging due to different factors that influence the imputation are data types, patterns, and the amount of data [15]. Deep learning can learn the hidden patterns and complex relations in data more effectively than traditional methods [16]. Generative adversarial networks can reconstruct the original data distribution, which is more suited for remote sensing data imputation [17].

In the literature, matrix factorization and deep learning methods [18, 19] can handle relationships within the data by recognizing the correct pattern, resulting in more imputations that are accurate. Deng et al. [20] imputed the values in the healthcare domain by mitigating biases and maintaining data integrity. Adversarial networks, such as Generative Adversarial Networks (GANs), are also used for imputation [21, 22]. GAN uses a generator network for imputing the missing data, and a discriminator is used to distinguish the observed value from the imputed value. This enforces the generator to produce more realistic values in the imputation process.

In natural language processing tasks, LLMs have shown better capabilities. In recent times, LLMs have utilized for data manipulation to improve data quality for downstream applications. Using LLM Prompting and fine-tuning methods in data error processing, the noisy data can be identified and corrected [23, 24]. Retrieval augmented generation (RAG) methods and contextual hints are used to impute missing values [25]. Peeters et al. used structured prompts [26] for imputing missing values. Nazir et al. [27] explored the usage of ChatGPT [28] as a data imputation method by prompting with text-based questions. Hayat and Hasan [29] introduced an approach that can generate text-specific descriptions for missing data. These methods need fine-tuning processes, which are computationally expensive. In context-based learning [30], the use of knowledge and reasoning [31] is used for tabular data classification, which avoids fine-tuning. Zhou et al. [32] introduced the LLM-based method for time series analysis tasks. RetClean [33] improved LLMs' performance by serializing each record into a format such as [Name: Johnroy; Gender: NULL; Age: 35] with a query such as "*what is the Gender value?*". Inspired by the ensemble learning, such as random forest, He et al. [34] proposed a framework called LLM Forest, where forest indicates a forest of few-shot learning LLM trees using a confidence-based weighted voting. Hayat et al. [35] proposed a novel approach, CRILM (contextually relevant imputation leveraging pre-trained language models for handling missing values in tabular data by creating contextually relevant descriptors for missing values. Yang et al. [36] propose an approach for retrieval augmented imputation (RAI) which utilize fine grained record (tuple) level retrieval rather than table based retrieval.

Large Language Models (LLMs) have demonstrated remarkable capabilities across a wide rangeof natural language processing tasks. UnIMP [37] frameworks have shown the ability of LLMs to discover the contextual relationships within table data, framing imputation methods as a fill-in-the-blank in the tables by using pre-training knowledge for filling the missing values based on the semantics of the data, along with patterns. Similar related work, such as TabLLM, also demonstrates the applications of LLMs to tabular data where these models understand and process the information that is in a structured format. There are a few successful applications of LLMs, such as text summarization and sentiment analysis [38], NER[39], and relation extraction [40] are present in the literature. Quantum-UnIMP [41] presented a novel framework that integrates shallow quantum circuits into a Large Language Model (LLM)-based imputation architecture to address missing data in mixed-type data scenarios.

## 3. Proposed Methodology

The primary objective of this research is to develop a computationally efficient data recovery framework suitable for low-resource environments. We propose a two-stage architecture that integrates density-based clustering with Retrieval-Augmented Generation (RAG). The following sections detail the granular steps of our two-stage approach. The proposed algorithm is presented in Algorithm 1 and is depicted in Figure 1.

### 3.1 Stage I: Representative Sample Extraction

In this stage, the objective is to reduce the search space from a massive raw dataset to a condensed set of "Effective Samples."

- **Step 1: Dataset Preprocessing.** The collected dataset undergoes all the necessary preprocessing steps such as normalization, one hot encoding etc. It is important to note that the given dataset is of heterogenous type and consists of both categorical and numerical attributes.
- **Step 2: Invoke Clustering algorithm** The pre-processed dataset obtained from the Step 1 undergoes with the Density-based spatial clustering of application with noise (DBSCAN) algorithm. This algorithm is particularly chosen based on its ability to identify clusters of arbitrary shapes and its inherent robustness against outliers.
- **Step 3: Employ Distance** Since we are handling the mixed dataset, we employ the Gower distance metric. This allows the subsequent clustering algorithm to treat qualitative and quantitative variables with equal mathematical weight.
- **Step 4: Identification of Cluster Centroids.** Upon convergence, suppose DBSCAN algorithm yields $K$ clusters. These centroids are considered to be the most representative sample for the entire cluster. Suppose there are $K$ clusters then $K$ number of centroids are generated.
- **Step 5: Collection of Local Effective Samples.** The collection of these $K$ centroids is formalized as the Local Effective Samples (LES). These represent the entire dataset covering the majority distribution of the data.
- **Step 6: Neighbourhood Aggregation and Global Synthesis.** To improve the representativeness and capture the variance around the centroids, we considered $t$ nearest neighbours for each centroid. The union of these neighbours with the corresponding LES constitutes the Global Effective Samples (GES). This ensures the local fluctuations does not lead to reduction in the quality of the dataset.

### 3.2 Stage II: Multi-LLM based Data Recovery

The second stage utilizes the condensed samples and determines the recommended value for a given missing sample through the multi-LLM RAG pipeline.

- **Step 7: RAG Indexing and Vector Store Integration.** Instead of querying the entire database, we populate a Retrieval-Augmented Generation (RAG) system with the LES and GES. This generates two different RAG i.e., (i) $RAG_{LES}$ and (ii) $RAG_{GES}$. This kind of settings significantly reduces the latency and enhances the token consumption to achieve our objective working with low-resource constraints.
- **Step 8: Specialized Model Deployment.** In our settings, we employed the following dual-model architecture:
    - $LLM_{local}$: Recommends the values by considering strictly by utilizing knowledge obtained from $RAG_{LES}$ ( LES samples).
    - $LLM_{global}$: Recommends the value by considering based on the broader $RAG_{GES}$ (GES samples neighbours).
- **Step 9: Parallel Inference and Contextual Retrieval.** When a record with a missing value is introduced, both $LLM_{local}$ and $LLM_{global}$ independently query the RAG system. They retrieve the most relevant samples and then recommends the appropriate value.
- **Step 10: Consensus-Based Voting Principle** To derive the final recommended value, we invoked the *Consensus Algorithm* by taking cue from the work of Sairam et al. The system evaluates the outputs of both LLMs, and then invokes the consensus algorithm to resolve discrepancies and recommend the final value.

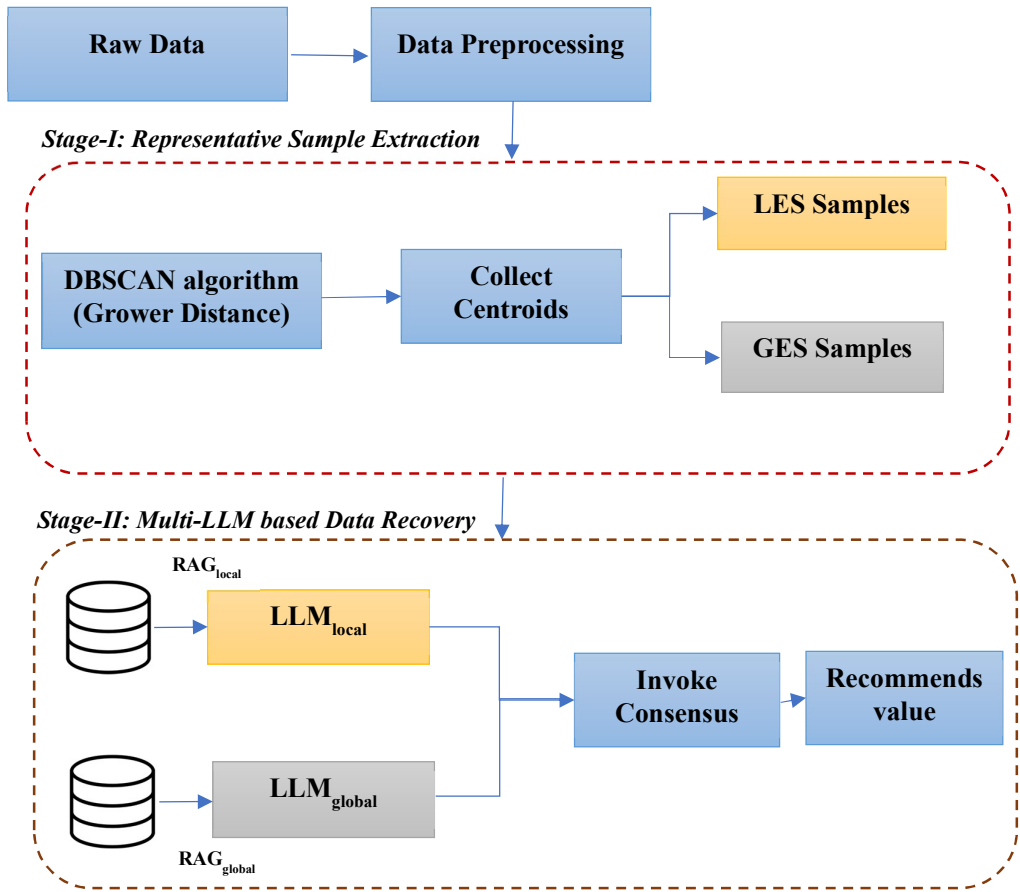

Fig. 1. Block diagram of the proposed LLMDR framework

**Algorithm 1: LLMDR procedure**

**Input:** Dataset $D$, parameters $\epsilon$, minPts, $t$, query record $x^*$

**Output:** Recommended value $V_{final}$

1. **Procedure** Stage-I $(D)$:
2. $D' \leftarrow$ Normalize and Encode$(D)$
3. $\{C_1, ..., C_K\} \leftarrow$ DBSCAN$(D', d_G, \epsilon, \text{minPts})$
4. $S_{LES} \leftarrow \setminus \{m_k \mid m_k \text{ is centroid of } C_k, \forall\, k \in [1, K]\}$
5. $S_{GES} \leftarrow S_{LES} \cup \{neighbours\; t\; of\; m_k\}$
6. **Return** $S_{LES}, S_{GES}$
7. **Procedure** Stage-II $(x^*, S_{LES}, S_{GES})$:
8. $R_{LES}, R_{GES} \leftarrow$ Construct RAG Indices$(S_{LES}, S_{GES})$
9. $v_{local} \leftarrow LLM_{local}.\text{predict}(x^*, R_{LES})$
10. $v_{global} \leftarrow LLM_{local}.\text{predict}(x^*, R_{LES})$
11. $V_{final} \leftarrow$ Consensus$(\{v_{local}, v_{global}\})$
12. **Return** $V_{final}$

## 4. Dataset Description

To evaluate the robustness of our LLM-driven consensus framework, we conducted experiments on three real-world datasets with distinct structural characteristics:

- **Buy Dataset**, an e-commerce collection of product metadata (Name, Description, and Manufacturer) that tests the model's ability to infer categorical data through semantic and brand associations;
- **Phone Dataset**, which represents a complex multi-type environment containing a mix of categorical labels and numerical values (Price, Rating, and Reviews) to assess the correlation between quantitative metrics and qualitative descriptors; and the
- **Restaurant Dataset**, a structured business directory (Name, Address, City, Phone, and Type) that serves as a benchmark for geographical and logical reasoning in recovering highly specific, non-prose identifiers.

All the datasets are available in the public repository.

## 4.1 Metrics

### 4.1.1. Accuracy

The number of exact predicitons, this is calculated for all the features. For example, the original value is 0.9 and the recovered value is turned out to be 0.9 then it is considered to be correct prediction otherwise not.

### 4.1.2 KS Complement Statistic

It measures the similarity of a given column in the real and synthetic datasets. It uses the KoglomorinovSmironov statistic (KS statistic), which is calculated by converting the numerical column into its cumulative distribution frequency (CDF). The maximum difference beten the two CDFs is known as the KS statistic, which lies between [0,1]. The KSComplement metric, mathematically defined in Eq. 6, also lies within the range of [0,1]. The higher the KSComplement the better the similarity is between real and synthetic columns.

### 4.1.3 SMAPE

SMAPE looks at the percentage of the error relative to the size of the values. It calculates the absolute difference between the prediction and the actual value, then divides it by the average of the two.

### 4.1.4 MSE

MSE calculates the average of the squares of the errors. In simple terms, it measures the variance between your predicted value and the actual value.

Table 1: LLM$_{local}$ performance metrics on various percentages of Missing values over BUY dataset

|  | Missing Rate (%) | Feature | Accuracy (%) | KS-Stat | SMAPE | MSE |
|---|---|---|---|---|---|---|
| LLM 1 | 10 | Description | 5.2632 | 0.1579 | 0.294 | 91.6316 |
|  |  | Manufacturer | 55 | 0.15 | 0.213 | 50.65 |
|  |  | Name | 0 | 0 | 1 | 130 |
|  | 20 | Description | 5.1282 | 0.3846 | 0.4955 | 826.7692 |
|  |  | Manufacturer | 35.8974 | 0.2308 | 0.3607 | 215.8718 |
|  |  | Name | 0 | – | 1 | 382.0256 |
|  | 30 | Description | 8.6207 | 0.431 | 0.4942 | 1180.603 |
|  |  | Manufacturer | 41.6667 | 0.3167 | 0.4318 | 315.6333 |
|  |  | Name | 0 | – | 1 | 696.431 |

## 5. Results & Discussion

In this section, we analyze the performance of the LLMDR and compared the performance with LLM$_{local}$ and LLM$_{global}$.

### 5.1 Analysis over the Buy Dataset

Upon the Buy Dataset which is having mixed types of data, we generated different percentages of missing values such as 10, 20 and 30% in MAR pattern. Now the LLM$_{local}$ is employed to recover the values by considering knowledge obtained from RAG$_{LES}$ ( LES samples). We can observe from Table 1 that the *Name* feature cannot be inferred due to 0% accuracy. *Description* feature increases the MSE indicating it is sensitive to the missing values. Where as *Manufacture* feature showed higher accuracy and lower error values even when the missing data increases. Other metrics such as SMAPE and MSE increases significantly indicating the performance degradation.

We also provided the mean, median and standard deviation values for various percentages of missing values in Table 2. For 10% of missing values we can observe highest mean accuracy with lowest error values. For MSE and KS-Stat the standard deviation is low which indicates a consistent behaviour across the features. On the other hand the large standard deviation in accuracy indicates some features are predicted much better compared to others. For 20% and 30% missing values, the mean values are low and degrade the performance. For SMAPE and MSE are increased, reflecting higher prediction errors. KS-Stat values shows higher difference between actual and predicted values as the percentage of missing values increases. The standard deviation is increased for all metrics, resulting the reduce in model performance.

**Table 2: Statistical Summary of $LLM_{local}$ on various percentages of missing values over BUY dataset**

|  | Missing Rate | Metric | Mean | Median | Standard Deviation |
|---|---|---|---|---|---|
| $LLM_{local}$ | 10% | Accuracy (%) | 20.0877 | 5.2632 | 30.3492 |
|  |  | KS-Stat | 0.1026 | 0.15 | 0.089 |
|  |  | SMAPE | 0.5023 | 0.294 | 0.4329 |
|  |  | MSE | 90.7605 | 91.6316 | 39.6822 |
|  | 20% | Accuracy (%) | 13.6752 | 5.1282 | 19.4151 |
|  |  | KS-Stat | 0.2051 | 0.2308 | 0.1936 |
|  |  | SMAPE | 0.6187 | 0.4955 | 0.337 |
|  |  | MSE | 474.8889 | 382.0256 | 315.8585 |
|  | 30% | Accuracy (%) | 16.7625 | 8.6207 | 21.9942 |
|  |  | KS-Stat | 0.2492 | 0.3167 | 0.2233 |
|  |  | SMAPE | 0.642 | 0.4942 | 0.3116 |
|  |  | MSE | 730.8893 | 696.431 | 433.5134 |

$LLM_{global}$ performance is presented in Tables 3 and 4. From Table 3, we can observe that the manufacture feature shown best performance in different percentages of missing values while maintaining low KS-Stat, MSE, and SMAPE values. That means $LLM_{global}$ can learn structured and categorical information that is related to manufacture feature. For the Description feature, as the percentage of missing values increases, MSE also increased indicating unstable predictions. Similarly, for the Name feature, observe very low accuracy and high KS-Stat and

SMAPE values due to high textual identifiers. LLM$_{global}$ can handle the structured features and is sensitive to the sparse or unstructured data.

**Table 3: LLM$_{global}$ performance metrics on various percentages of Missing values over BUY dataset**

|  | Missing Rate (%) | Feature | Accuracy (%) | KS-Stat | SMAPE | MSE |
|---|---|---|---|---|---|---|
| LLM$_{global}$ | 10 | Description | 5.2632 | 0.1579 | 0.294 | 91.6316 |
|  |  | Manufacturer | 55 | 0.15 | 0.213 | 50.65 |
|  |  | Name | 0 | 0 | 1 | 130 |
|  | 20 | Description | 10.2564 | 0.17949 | 0.23497 | 491.333 |
|  |  | Manufacturer | 64.1026 | 0.17949 | 0.11139 | 48.1026 |
|  |  | Name | 2.5641 | 0.28205 | 0.15462 | 207.179 |
|  | 30 | Description | 10.3448 | 0.24138 | 0.2544 | 1395.36 |
|  |  | Manufacturer | 65 | 0.15 | 0.10544 | 81.55 |
|  |  | Name | 1.72414 | 0.25862 | 0.16349 | 913.086 |

**Table 4: Statistical Summary of LLM$_{global}$ on various percentages of missing values over BUY dataset**

|  | Missing Rate | Metric | Mean | Median | Standard Deviation |
|---|---|---|---|---|---|
| LLM$_{global}$ | **10%** | Accuracy (%) | **26.93** | **10.53** | **33.17** |
|  |  | KS-Stat | **0.1737** | **0.1579** | **0.0831** |
|  |  | SMAPE | **0.1776** | **0.1765** | **0.0986** |
|  |  | MSE | **72.78** | **65.16** | **74.43** |
|  | **20%** | Accuracy (%) | **25.64** | **10.26** | **35.1** |
|  |  | KS-Stat | **0.2137** | **0.1795** | **0.0588** |
|  |  | SMAPE | **0.167** | **0.1546** | **0.0622** |
|  |  | MSE | **248.87** | **207.18** | **224.27** |
|  | **30%** | Accuracy (%) | **25.69** | **10.34** | **36.25** |
|  |  | KS-Stat | **0.216** | **0.2414** | **0.0567** |
|  |  | SMAPE | **0.1744** | **0.1635** | **0.0758** |
|  |  | MSE | **796.67** | **913.09** | **659.87** |

Table 4 shows the statistical results of $LLM_{global}$. For various percentages of missing values, the accuracy remains relatively stable. However, substantial variability in performance between mean, median, and standard deviation. $LLM_{global}$ performs inconsistently as the number of missing values increases, exhibiting degradation in prediction reliability. From the KS-Stat and SMAPE values, we can observe that a higher relative error is associated with reduced distributional similarity. The MSE and standard deviation values indicate that it is also sensitive to the missing information.

To derive the final recommended value, we invoked the Consensus Algorithm (LLMDR). We evaluate the outputs of both LLMs and then invoke the consensus algorithm to resolve discrepancies and recommend the final value with a stabilized and aggregated value. Table 5 shows that manufacture feature achieves higher accuracy for missing rates by maintaining low KS-Stat and MSE values. For description and Name features, shown weaker performance under higher percentages of missing values. Name features have almost zero accuracy, with an increasing error metric for higher missing values. The consensus predictions for the manufacture feature are more reliable than the name and description features.

**Table 5: LLMDR performance metrics on various percentages of Missing values over BUY dataset**

|  | Missing Rate (%) | Feature | Accuracy (%) | KS-Stat | SMAPE | MSE |
|---|---|---|---|---|---|---|
| LLMDR | 10 | Description | 5.26316 | 0.15789 | 0.29732 | 183.737 |
|  |  | Manufacturer | 70 | 0.05 | 0.06972 | 0.3 |
|  |  | Name | 0 | 0.10526 | 0.1015 | 2.15789 |
|  | 20 | Description | 10.2564 | 0.15385 | 0.23783 | 393.333 |
|  |  | Manufacturer | 69.2308 | 0.10256 | 0.04791 | 7.35897 |
|  |  | Name | 2.5641 | 0.12821 | 0.1214 | 124.667 |
|  | 30 | Description | 10.3448 | 22.4138 | 25.6699 | 1429.9 |
|  |  | Manufacturer | 76.6667 | 6.66667 | 6.69189 | 31.65 |
|  |  | Name | 0 | 15.5172 | 14.2205 | 856.31 |

Table 6 shows the statistical results of LLMDR, which show a trade-off between accuracy and error stability as the percentage of missing values increases. Mean accuracy for various percentages of missing values indicates that Consensus extracts useful patterns even with higher data loss. KS-Stat and SMAPE remain relatively stable in prediction error. For a higher percentage of missing values (30%), there is a sharp rise in KS-Stat, MSE, and SMAPE.

**Table 4: Statistical Summary of LLMDR on various percentages of missing values over BUY dataset**

| | Missing Rate | Metric | Mean | Median | Standard Deviation |
|---|---|---|---|---|---|
| LLMDR | 10% | Accuracy (%) | 25.0877 | 5.2632 | 38.9841 |
| | | KS-Stat | 0.1044 | 0.1053 | 0.054 |
| | | SMAPE | 0.1562 | 0.1015 | 0.1233 |
| | | MSE | 62.0649 | 2.1579 | 105.3751 |
| | 20% | Accuracy (%) | 27.3504 | 10.2564 | 29.7799 |
| | | KS-Stat | 0.1282 | 0.1282 | 0.0209 |
| | | SMAPE | 0.1357 | 0.1214 | 0.0782 |
| | | MSE | 175.1197 | 124.6667 | 161.5615 |
| | 30% | Accuracy (%) | 29.0038 | 10.3448 | 33.9663 |
| | | KS-Stat | 14.8659 | 15.5172 | 6.4452 |
| | | SMAPE | 15.5274 | 14.2205 | 7.8027 |
| | | MSE | 772.619 | 856.3103 | 573.8911 |

## 5.2 Analysis over the Phone Dataset

For the phone dataset, the results of $LLM_{local}$, $LLM_{global}$, and LLMDR, along with statistical results, are given in tables 7-12.

Table 7 shows the performance of $LLM_{local}$ for various missing rates. Brand name is the most robust feature as we can observe that the accruacy steadily improves as missingness increases with strong KS values. For price and product name features has error prone predictions for more than 20% interms of KS, SMAPE and MSE. As missingness increases Rating maintains moderate accruacy with increasing SMAPE and low MSE. Noisy predictions observed in review votes feature with moderate accuracy and high SMAPE. The results indicate that textual or categorical features such as brand name generalizes best, whereas numeric, sensitive features such as price and reviews are highly vulnerable to missing data.

From Table 8, $LLM_{global}$ has shown better performance on structured features. compared to other features rating features exhibits high accuracy with low MSE and SMAPE values. Brand name feature also scales well with missingness with improved accuracy. Price accuracy improves at 30% but price and product features shown rising MSE. The relative error control happened for review votes feature. among all the features reviews is near zero accuracy and high MSE value for higher percentage of missing values. Table 9 show LLMDR results which are more balanced performance under increasing missingness. Brand name feature contain improved accuracy from 30% to 48% while KS predictably declines as missingness increases. Rating features maintain high accuracy with very low MSE and SMAPE values. Among non-rating features review votes steadily improves with missingness. Price and product name accuracy varies. MSE value increases with increasing missingness as its sensitivity to sparsity and scale. reviews feature have zero accuracy and escalating MSE for various percentages of missing values.

**Table 7: LLM$_{local}$ performance metrics on various percentages of Missing values over Phone dataset**

|  | Missing Rate (%) | Feature | Accuracy (%) | KS-Stat | SMAPE | MSE |
|---|---|---|---|---|---|---|
| LLM$_{local}$ | 10 | Brand Name | 26.74419 | 0.651163 | 0.458201 | 48.94186 |
|  |  | Price | 11.11111 | 0.166667 | 0.31327 | 318.3056 |
|  |  | Product Name | 0 | 0.266667 | 0.307109 | 357.7333 |
|  |  | Rating | 46.66667 | 0.2 | 0.297926 | 11.6 |
|  |  | Review Votes | 41.46341 | 0.341463 | 0.481424 | 37.41463 |
|  |  | Reviews | 0 | 0.3 | 0.366979 | 395.8667 |
|  | 20 | Brand Name | 39.65517 | 0.482759 | 0.267848 | 35.26724 |
|  |  | Price | 12.12121 | 0.166667 | 0.315145 | 827.197 |
|  |  | Product Name | 8.333333 | 0.166667 | 0.300581 | 806.5833 |
|  |  | Rating | 51.66667 | 0.233333 | 0.340827 | 13.41667 |
|  |  | Review Votes | 39.43662 | 0.169014 | 0.521792 | 69.22535 |
|  |  | Reviews | 0 | 0.316667 | 0.441457 | 2232.683 |
|  | 30 | Brand Name | 45.89041 | 0.383562 | 0.240571 | 51.70548 |
|  |  | Price | 16.66667 | 0.15625 | 0.335002 | 1712.958 |
|  |  | Product Name | 6.666667 | 0.077778 | 0.223857 | 980.7222 |
|  |  | Rating | 40 | 0.133333 | 0.318567 | 17.27778 |
|  |  | Review Votes | 39.60396 | 0.227723 | 0.515649 | 46.74257 |
|  |  | Reviews | 0 | 0.177778 | 0.414199 | 4742.744 |

**Table 8: LLM$_{global}$ performance metrics on various percentages of Missing values over Phone dataset**

| | Missing Rate (%) | Feature | Accuracy (%) | KS-Stat | SMAPE | MSE |
|---|---|---|---|---|---|---|
| LLM$_{global}$ | 10 | Brand Name | 29.06977 | 0.651163 | 0.667268 | 9.72093 |
| | | Price | 27.77778 | 0.277778 | 0.296131 | 280.75 |
| | | Product Name | 23.33333 | 0.2 | 0.2285 | 122.9667 |
| | | Rating | 83.33333 | 0.033333 | 0.073968 | 1.033333 |
| | | Review Votes | 34.14634 | 0.268293 | 0.442055 | 6.195122 |
| | | Reviews | 0 | 0.2 | 0.418757 | 612.1333 |
| | 20 | Brand Name | 48.27586 | 0.491379 | 0.490016 | 14.57759 |
| | | Price | 24.24242 | 0.227273 | 0.320595 | 931.1515 |
| | | Product Name | 23.33333 | 0.25 | 0.282142 | 450.0667 |
| | | Rating | 81.66667 | 0.05 | 0.109365 | 1.45 |
| | | Review Votes | 32.39437 | 0.338028 | 0.425758 | 19.47887 |
| | | Reviews | 0 | 0.2 | 0.414513 | 2381.133 |
| | 30 | Brand Name | 50 | 0.390411 | 0.413893 | 9.041096 |
| | | Price | 37.5 | 0.177083 | 0.233972 | 1024.333 |
| | | Product Name | 15.55556 | 0.211111 | 0.273884 | 1182.856 |
| | | Rating | 66.66667 | 0.088889 | 0.15512 | 3.811111 |
| | | Review Votes | 43.56436 | 0.29703 | 0.335607 | 16.85149 |
| | | Reviews | 1.111111 | 0.155556 | 0.385827 | 4904.678 |

**Table 9: LLMDR performance metrics on various percentages of Missing values over Phone dataset**

| | Missing Rate (%) | Feature | Accuracy (%) | KS-Stat | SMAPE | MSE |
|---|---|---|---|---|---|---|
| LLMDR | 10 | Brand Name | 30.23256 | 0.651163 | 0.421759 | 38.33721 |
| | | Price | 30.55556 | 0.305556 | 0.222725 | 200.1944 |
| | | Product Name | 23.33333 | 0.1 | 0.171445 | 71.76667 |
| | | Rating | 73.33333 | 0.066667 | 0.112686 | 1.633333 |
| | | Review Votes | 39.02439 | 0.365854 | 0.534553 | 19.29268 |
| | | Reviews | 0 | 0.2 | 0.392715 | 507.8 |
| | 20 | Brand Name | 47.41379 | 0.482759 | 0.241092 | 27.7069 |
| | | Price | 18.18182 | 0.212121 | 0.246202 | 602.9242 |
| | | Product Name | 23.33333 | 0.1 | 0.258106 | 532.5167 |
| | | Rating | 66.66667 | 0.116667 | 0.170226 | 2.483333 |
| | | Review Votes | 43.66197 | 0.197183 | 0.538473 | 20.47887 |
| | | Reviews | 0 | 0.216667 | 0.412583 | 2694.55 |
| | 30 | Brand Name | 47.94521 | 0.390411 | 0.276335 | 37.19178 |
| | | Price | 29.16667 | 0.125 | 0.243464 | 991.4271 |
| | | Product Name | 13.33333 | 0.2 | 0.233659 | 861.0778 |
| | | Rating | 65.55556 | 0.1 | 0.144535 | 3.111111 |
| | | Review Votes | 47.52475 | 0.188119 | 0.48906 | 21.19802 |
| | | Reviews | 0 | 0.2 | 0.367693 | 4821.111 |

Statistical results of the Phone dataset for $LLM_{local}$, $LLM_{global}$, and LLMDR are presented in tables 10 to 12. Across increasing missing rates (see table 10), $LLM_{local}$ mean accuracy improves for a higher percent of missing rate but the standard deviation value indicates the high variability across features. Reduced class separation under high missingness is noticed for KS-Stat. Also, SMAPE decreases for 30% missing values, and MSE exponentially grows with very high dispersion. $LLM_{global}$ showed consistently better performance than $LLM_{local}$ as presented in Table 11, particularly in accuracy, even with a high percentage of missing values. It indicates the improved robustness for certain features. Similar to $LLM_{local}$, KS-Stat values are low but remain competitive. Also, better relative error control is observed with respect to SMAPE, as it improves significantly with higher missing rates, and MSE exhibits very high variance. This indicates that $LLM_{local}$ handles proportional errors better than absolute prediction errors. Compared to both $LLM_{local}$ and $LLM_{global}$, we can observe that $LLM_{global}$ offers steady accuracy with missing rates. From Table 12, LLMDR achieves the lowest SMAPE means across all missing rates, despite increasing variance. It's also maintained a lower MSE than $LLM_{local}$ and $LLM_{global}$. LLMDR exhibits the most robustness with balancing accuracy, stability, and error control.

**Table 10: Statistical Summary of LLM_local on various percentages of missing values over Phone dataset**

| | Missing Rate | Metric | Mean | Median | Standard Deviation |
|---|---|---|---|---|---|
| LLM_local | 10% | Accuracy (%) | 20.9976 | 18.9276 | 20.4466 |
| | | KS-Stat | 0.321 | 0.2833 | 0.1739 |
| | | SMAPE | 0.3708 | 0.3401 | 0.0807 |
| | | MSE | 194.977 | 183.6237 | 179.9086 |
| | 20% | Accuracy (%) | 25.2022 | 25.7789 | 20.989 |
| | | KS-Stat | 0.2559 | 0.2012 | 0.1258 |
| | | SMAPE | 0.3646 | 0.328 | 0.097 |
| | | MSE | 664.0621 | 437.9043 | 857.9104 |
| | 30% | Accuracy (%) | 24.8046 | 28.1353 | 19.5193 |
| | | KS-Stat | 0.1927 | 0.167 | 0.1058 |
| | | SMAPE | 0.3413 | 0.3268 | 0.1097 |
| | | MSE | 1258.692 | 516.2139 | 1837.879 |

**Table 11: Statistical Summary of LLM_global on various percentages of missing values over Phone dataset**

| | Missing Rate | Metric | Mean | Median | Standard Deviation |
|---|---|---|---|---|---|
| LLM_global | 10% | Accuracy (%) | 32.94343 | 28.42377 | 27.42226 |
| | | KS-Stat | 0.271761 | 0.234146 | 0.205495 |
| | | SMAPE | 0.354447 | 0.357444 | 0.203764 |
| | | MSE | 172.1332 | 66.3438 | 241.2482 |
| | 20% | Accuracy (%) | 34.98544 | 28.3184 | 27.70434 |
| | | KS-Stat | 0.259447 | 0.238636 | 0.147305 |
| | | SMAPE | 0.340398 | 0.367554 | 0.135888 |
| | | MSE | 632.9763 | 234.7728 | 931.2399 |
| | 30% | Accuracy (%) | 35.73295 | 40.53218 | 23.79368 |
| | | KS-Stat | 0.220013 | 0.194097 | 0.107924 |
| | | SMAPE | 0.299717 | 0.304745 | 0.097609 |
| | | MSE | 1190.262 | 520.5924 | 1897.594 |

**Table 12: Statistical Summary of LLMDR on various percentages of missing values over Phone dataset**

| | Missing Rate | Metric | Mean | Median | Standard Deviation |
|---|---|---|---|---|---|
| LLMDR | 10% | Accuracy (%) | 32.74653 | 30.39406 | 23.91135 |
| | | KS-Stat | 0.28154 | 0.252778 | 0.214515 |
| | | SMAPE | 0.309314 | 0.30772 | 0.164621 |
| | | MSE | 139.8374 | 55.05194 | 193.7043 |
| | 20% | Accuracy (%) | 33.2096 | 33.49765 | 23.90246 |
| | | KS-Stat | 0.220899 | 0.204652 | 0.137629 |
| | | SMAPE | 0.311114 | 0.252154 | 0.136941 |
| | | MSE | 646.7767 | 280.1118 | 1039.131 |
| | 30% | Accuracy (%) | 33.92092 | 38.34571 | 24.41137 |
| | | KS-Stat | 0.200588 | 0.194059 | 0.102 |
| | | SMAPE | 0.292457 | 0.259899 | 0.120259 |
| | | MSE | 1122.519 | 449.1348 | 1865.96 |

### 5.3 Analysis over the Restaurant Dataset

Table 13 to 15 describes the results of the $LLM_{local}$, $LLM_{global}$ and LLMDR. From Table 13, we can observe that $LLM_{local}$ performance is affected by feature type and missing rate. For 10 and 30% of missing values, the model performs best on features such as city compared to phone, address, and name, which show low accuracy and high error. Under high data loss, we can observe that error metrics (SMAPE and MSE) are increased for address, name, and phone features. The type feature exhibits consistently high relative error. Table 14 shows the performance of $LLM_{global}$, particularly for features such as city, which achieved higher accuracy across all missing rates with low KS, SMAPE, and MSE values. The feature type show high SMAPE and moderate accuracy. Other attributes, such as address and name exhibits instability. For LLMDR, performance is presented in Table 15. For categorical features such as city, LLMDR exhibits consistent performance and low KS, SMAPE, and MSE values. The feature called type LLMDR remains moderately accurate even for higher missing rates with increasing of SMAPE. Especially for 30% missing values, features such as address, name, and phone exhibit low accuracy and increasing of MSE value.

**Table 13: LLM$_{local}$ performance metrics on various percentages of Missing values over Restaurant dataset**

|  | Missing Rate (%) | Feature | Accuracy (%) | KS-Stat | SMAPE | MSE |
|---|---|---|---|---|---|---|
| LLM$_{local}$ | 10 | addr | 0 | 0.192308 | 0.381645 | 443.2308 |
|  |  | city | 26.92308 | 0.115385 | 0.346242 | 22.30769 |
|  |  | name | 7.692308 | 0.461538 | 0.448194 | 343.6923 |
|  |  | phone | 0 | 0.192308 | 0.254548 | 131.0769 |
|  |  | type | 7.692308 | 0.346154 | 0.493281 | 160.7692 |
|  | 20 | addr | 0 | 0 | 0 | 0 |
|  |  | city | 42.30769 | 0 | 0 | 0 |
|  |  | name | 3.846154 | 0 | 0 | 0 |
|  |  | phone | 0 | 0 | 0 | 0 |
|  |  | type | 3.846154 | 0 | 0 | 0 |
|  | 30 | addr | 0 | 0.217949 | 0.386911 | 3762.756 |
|  |  | city | 32.05128 | 0.192308 | 0.265156 | 30.89744 |
|  |  | name | 0 | 0.153846 | 0.357546 | 2874.167 |
|  |  | phone | 0 | 0.217949 | 0.247701 | 1561.09 |
|  |  | type | 6.410256 | 0.102564 | 0.523821 | 351.5128 |

**Table 14: LLM$_{global}$ performance metrics on various percentages of Missing values over Restaurant dataset**

|  | Missing Rate (%) | Feature | Accuracy (%) | KS-Stat | SMAPE | MSE |
|---|---|---|---|---|---|---|
| LLM$_{global}$ | 10 | addr | 11.53846 | 0.269231 | 0.344482 | 326 |
|  |  | city | 69.23077 | 0.038462 | 0.120706 | 5.538462 |
|  |  | name | 7.692308 | 0.461538 | 0.420473 | 440.5 |
|  |  | phone | 0 | 0.153846 | 0.181179 | 33.84615 |
|  |  | type | 34.61538 | 0.423077 | 0.436591 | 101.4615 |
|  | 20 | addr | 9.615385 | 0.211538 | 0.281442 | 1092.519 |
|  |  | city | 71.15385 | 0.057692 | 0.090838 | 6.25 |
|  |  | name | 11.53846 | 0.192308 | 0.352327 | 1430.269 |
|  |  | phone | 3.846154 | 0.076923 | 0.162246 | 172.9038 |
|  |  | type | 17.30769 | 0.192308 | 0.406471 | 149.6154 |
|  | 30 | addr | 8.974359 | 0.166667 | 0.32912 | 3087.308 |
|  |  | city | 64.10256 | 0.128205 | 0.104199 | 21.25641 |
|  |  | name | 6.410256 | 0.205128 | 0.363215 | 3123.308 |
|  |  | phone | 2.564103 | 0.102564 | 0.210772 | 762.1795 |
|  |  | type | 30.76923 | 0.230769 | 0.406236 | 214.5513 |

**Table 15: LLMDR performance metrics on various percentages of Missing values over Restaurant dataset**

|       | Missing Rate (%) | Feature | Accuracy (%) | KS-Stat  | SMAPE    | MSE      |
|-------|------------------|---------|--------------|----------|----------|----------|
| LLMDR | 10               | addr    | 11.53846     | 0.192308 | 0.336195 | 363.7692 |
|       |                  | city    | 61.53846     | 0.076923 | 0.133531 | 8.153846 |
|       |                  | name    | 3.846154     | 0.769231 | 0.479708 | 549.0769 |
|       |                  | phone   | 0            | 0.115385 | 0.207349 | 51       |
|       |                  | type    | 46.15385     | 0.269231 | 0.299354 | 61.92308 |
|       | 20               | addr    | 7.692308     | 0.153846 | 0.301486 | 1123.673 |
|       |                  | city    | 65.38462     | 0.096154 | 0.124104 | 11.69231 |
|       |                  | name    | 3.846154     | 0.557692 | 0.438693 | 1987.115 |
|       |                  | phone   | 3.846154     | 0.076923 | 0.176645 | 153.8654 |
|       |                  | type    | 21.15385     | 0.173077 | 0.460777 | 120.2692 |
|       | 30               | addr    | 7.692308     | 0.115385 | 0.349363 | 3374.346 |
|       |                  | city    | 61.53846     | 0.128205 | 0.13868  | 25.60256 |
|       |                  | name    | 7.692308     | 0.320513 | 0.384117 | 3326.667 |
|       |                  | phone   | 2.564103     | 0.141026 | 0.21447  | 746.0256 |
|       |                  | type    | 33.33333     | 0.217949 | 0.384433 | 248.5385 |

LLM$_{local}$ shows a moderate accuracy with low SMAPE and MSE values at 10% of missing values (See Table 16). At 20% the zero values across SMAPE, KS Stat, and MSE indicate that the model fails to generate meaningful outputs. When the missing rate increases to 30%, the model shows instability and increasing of error. Stable performance of LLM$_{global}$ is shown in Table 17. For a lower percentage of missing values, the mean accuracy is higher with a large standard deviation, indicating a variability across features. For 20 and 30% of missing values, the mean slightly decreases, but it will remain consistent. KS-Stat values remain moderate and show reduced variability for higher missing values. SMAPE values are low across all levels of missing rates. Table 18 shows the consensus results on Restaurant dataset. For 10% missing data model, the mean accuracy of 24% with a high standard deviation, which indicates the variability in features. For 20 and 30%, the mean accuracy slightly decreases, and performance is stable even for high percentages of missing values.

**Table 16: Statistical Summary of LLM$_{local}$ on various percentages of missing values over Restaurant dataset**

|  | Missing Rate | Metric | Mean | Median | Standard Deviation |
|---|---|---|---|---|---|
| LLM$_{local}$ | 10% | Accuracy (%) | 8.461538 | 7.692308 | 11.01371 |
|  |  | KS-Stat | 0.261538 | 0.192308 | 0.139738 |
|  |  | SMAPE | 0.384782 | 0.381645 | 0.092532 |
|  |  | MSE | 220.2154 | 160.7692 | 170.0216 |
|  | 20% | Accuracy (%) | 10 | 3.846154 | 18.16264 |
|  |  | KS-Stat | 0 | 0 | 0 |
|  |  | SMAPE | 0 | 0 | 0 |
|  |  | MSE | 0 | 0 | 0 |
|  | 30% | Accuracy (%) | 7.692308 | 0 | 13.8971 |
|  |  | KS-Stat | 0.176923 | 0.192308 | 0.049155 |
|  |  | SMAPE | 0.356227 | 0.357546 | 0.110796 |
|  |  | MSE | 1716.085 | 1561.09 | 1601.246 |

**Table 17: Statistical Summary of LLM$_{global}$ on various percentages of missing values over Restaurant dataset**

|  | Missing Rate | Metric | Mean | Median | Standard Deviation |
|---|---|---|---|---|---|
| LLM$_{global}$ | 10% | Accuracy (%) | 24.61538 | 11.53846 | 28.07956 |
|  |  | KS-Stat | 0.269231 | 0.269231 | 0.178339 |
|  |  | SMAPE | 0.300686 | 0.344482 | 0.142664 |
|  |  | MSE | 181.4692 | 101.4615 | 191.7886 |
|  | 20% | Accuracy (%) | 22.69231 | 11.53846 | 27.51412 |
|  |  | KS-Stat | 0.146154 | 0.192308 | 0.072722 |
|  |  | SMAPE | 0.258665 | 0.281442 | 0.130958 |
|  |  | MSE | 570.3115 | 172.9038 | 645.2348 |
|  | 30% | Accuracy (%) | 22.5641 | 8.974359 | 25.68266 |
|  |  | KS-Stat | 0.166667 | 0.166667 | 0.05286 |
|  |  | SMAPE | 0.282709 | 0.32912 | 0.123425 |
|  |  | MSE | 1441.721 | 762.1795 | 1542.817 |

**Table 18: Statistical Summary of LLM$_{global}$ on various percentages of missing values over Restaurant dataset**

| | Missing Rate | Metric | Mean | Median | Standard Deviation |
|---|---|---|---|---|---|
| LLMDR | 10% | Accuracy (%) | 24.61538 | 11.53846 | 27.5477 |
| | | KS-Stat | 0.284615 | 0.192308 | 0.280796 |
| | | SMAPE | 0.291227 | 0.299354 | 0.13183 |
| | | MSE | 206.7846 | 61.92308 | 237.9685 |
| | 20% | Accuracy (%) | 20.38462 | 7.692308 | 26.14253 |
| | | KS-Stat | 0.211538 | 0.153846 | 0.197525 |
| | | SMAPE | 0.300341 | 0.301486 | 0.151032 |
| | | MSE | 679.3231 | 153.8654 | 857.6342 |
| | 30% | Accuracy (%) | 22.5641 | 7.692308 | 24.88633 |
| | | KS-Stat | 0.184615 | 0.141026 | 0.085811 |
| | | SMAPE | 0.294213 | 0.349363 | 0.111595 |
| | | MSE | 1544.236 | 746.0256 | 1669.475 |

In summary, the results on three datasets show that LLM consensus has an advantage as it combines the recommended values by multiple LLMs instead of depending on a single model output. This consensus approach helps in reducing the bias of individual models, especially with higher percentages of missing values. The final output reflects a shared agreement across the models. The LLM consensus improves the confidence in decision-making while dealing with various percentages of missing values in mixed data types.

# 6. Conclusions

In this work, we introduced an LLM consensus framework for recovering various percentages of missing values in mixed data. LLMDR utilizes two LLMs called LLM local and LLM global, which recommend the value considering knowledge obtained from LES samples and GES samples. The final recommendation is given by the LLM Consensus. All three LLMs show the ability to recover with reasonable accuracy. Among them, LL2 achieves higher accuracy with lower error values for numerical and structured attributes. LLM shows moderate performance for categorical features, while stable performance is exhibited by LLMDR. In the future, we extend this work by addressing the level images by employing more advanced models of AI.